\documentclass[12pt]{article}

\usepackage{graphicx,color}
\usepackage{amssymb}
\usepackage{amsmath}

\numberwithin{equation}{section}

\begin{document}
\baselineskip=16pt
\begin{titlepage}

\begin{center}
\vspace*{12mm}

{\Large\bf%
Membrane scattering from Bagger-Lambert theory
}\vspace*{10mm}

Takayuki Hirayama$^{1,2}$\footnote{e-mail: 
{\tt hirayama@phys.cts.nthu.edu.tw}}
and
Dan Tomino$^{1}$\footnote{e-mail: 
{\tt tomino@phys.cts.nthu.edu.tw}}
\vspace*{5mm}

${}^1$
{\it Physics Division, National Center for Theoretical Sciences, Hsinchu 300, Taiwan 
\\[3mm]
${}^2$
Department of Physics, National Taiwan Normal University, Taipei 116, Taiwan
}\\[1mm]

\end{center}
\vspace*{10mm}

\begin{abstract}\noindent%
In this note, we discuss membrane scattering from the three dimensional $N=8$ superconformal theory with $SO(8)$ global symmetry constructed by Bagger-Lambert and Gustavsson. We discuss whether the one loop effective potential consistently reproduces the Newton potential of membranes moving in an eleven dimensional orbifold space.
\end{abstract}

\end{titlepage}

\newpage

\section{Introduction}

Through the understanding of D-branes and string duality (see, for example, the text book by Polchinski~\cite{Pol}), a more fundamental underling theory called M-theory has been expected and each string theory is realized as various limits in M-theory. Although this M-theory is expected, we have only poor understanding. Its low energy effective theory is given by the eleven dimensional supergravity and it would be a theory of membrane compared from a string theory which is a theory of string. The strong string coupling limit of IIA string theory opens up the eleventh space and is described by M-theory. It is conjectured that M-theory in an infinite momentum frame is described by BFSS matrix model~\cite{Banks:1996vh}.

The quantization of a membrane worldvolume theory is very challenging
and one of difficulty is the nonlocality associated with the deformation
of membrane without changing its volume (see, for example, a review by Taylor~\cite{tay}). In string theory, the open
string and closed string duality appears in many situations and has
provided many powerful techniques. One important idea behind BFSS matrix
model is also based on the open-closed string duality and the worldvolume theory of
multiple D0-brane, (which is governed by open string fluctuations) describes the target space dynamics, i.e. the gravity in the target space (which is governed by closed string fluctuations). Therefore another
direction to approach to M-theory is studying the effective action for
multiple Membrane.

Recently Bagger and Lambert (BL) constructed a new three dimensional N=8
superconformal theory using a three algebra~\cite{Bagger:2006sk} (see also~\cite{Gustavsson:2007vu} by Gustavsson). Since BL theory
satisfies all the properties which multiple membrane should have, it is
expected to describe multiple membranes. For BL theory with $SO(4)$ gauge symmetry, the moduli space~\cite{Bagger:2007vi, Lambert:2008et, Distler:2008mk} is discussed and the theory is conjectured to describe a two membrane system in an orbifold space~\cite{Lambert:2008et, Distler:2008mk}. Soon after the work
by Bagger-Lambert, Aharony, Bergman, Jafferis and Maldacena (ABJM) generalize their idea and constructed three
dimensional N=6 superconformal theories which contain BL theory as a special case~\cite{Aharony:2008ug}. ABJM also show the membrane configuration in
the eleven dimensional orbifold space time ($R^{1,2}\times (R^8/Z_k)$ and $k$ is the
level of Chern-Simons coupling) for their N=6 theory
with $U(N)\times U(N)$ gauge symmetry. Since the matter fields are charged under $U(1)$ in $U(N)=U(1)\times SU(N)$ and then $U(1)$ is not decoupled from $SU(N)$ in ABJM theory, BL theory with $SO(4)=SU(2)\times SU(2)$ may not
describe a multiple membrane system. However $U(1)$ gauge coupling is IR
free and the BL theory and ABJM theory with $U(2)\times U(2)$ gauge symmetry
may be connected by a renormalization flow. The target space superalgebra is studied in BL theory with the central element which suggests the target space is an eleven dimensional space~\cite{Passerini:2008qt}. It is also discussed that BL theory with the Nambu-Poisson algebra turns out be an action of single M5-brane~\cite{Ho:2008nn}. Therefore we may still
expect that BL theory describes multiple membranes. If so,
it worths studying a possibility that multiple membrane dynamics can describe
a target space dynamics, as parallel to that the D-brane dynamics describes the target space dynamics.

One important consequence of open-closed string duality is probing the
target space from D-brane scattering using D-brane effective theory,
i.e. Super Yang-Mills theory (SYM)~\cite{Douglas:1996yp} where the one loop effective potential reproduces the Newton potential in the target space. We then expect a similar correspondence in M-theory, and in
this note we study the one loop effective potential around a  relatively
moving membrane background in BL theory and see if the potential is understood as the Newton potential in the target space. Since the target space is discussed to be an orbifold $R^{1,2}\times (R^8/Z_k)$, we study a small $k$ case in order to probe the whole spacetime otherwise the one
spacial direction is effectively compactified in a large $k$
case ($Z_k$ is a subgroup of a $U(1)$ and we can always define one spacial direction generated by this $U(1)$ for any value of $k$. We call this direction the compactified direction.).
However the coupling constant is proportional to $1/k$, the theory
is strongly coupled for a small $k$ and the perturbation will not be a
good expansion. Despite of that we still expect the one loop effective
potential qualitatively gives a correct answer, since we expect that an
one loop open membrane amplitude can be reinterpreted as a tree closed
membrane amplitude, and also we treat a small deviation from BPS state. This situation is similar to BFSS matrix model. One should take a large N limit (N is the size of matrix) to recover the eleven dimensional Lorentz invariance, and the matrix model should give a controllable description at a shorter distance than the Plank 
length~\cite{Douglas:1996yp} at which we may expect the spacetime no longer looks like a eleven (or ten) dimensional classical spacetime. Despite of these, even for a finite N (N is the size of matrix), the one loop effective potential reproduces the Newton potential.

With this expectation in mind, we study the membrane scattering and compute the one loop effective potential in BL theory. The membrane scattering in ABJM theory is mentioned in~\cite{Berenstein:2008dc} and that in Lorentzian BL theory is discussed in~\cite{Verlinde:2008di}. From our calculations, we find that the potential is understood as the Newton potential and the total dimension of target space, which is read from the exponent of the power law behavior, is ten. The potential does not show a different behavior depending on the value of $k$. Thus in a large $k$ limit, the potential consistently becomes the one computed from D2-brane SYM action. This result suggests that the open membrane, described as a perturbation from the background, always wraps the compactified direction even $k$ is finite and small, and the BL theory can probe only
remaining ten dimensions. 

Using the three dimensional SYM action for multiple
D2-branes, the membrane scattering has been discussed. 
Polchinski and Pouliot discussed the
membrane scattering with momentum transfer along the eleventh
direction (M-momentum transfer) corresponds to an instanton process~\cite{Polchinski:1997pz}. We will see that the same happens in BL theory.

In the next section, we review the moduli space of BL theory and
introduce the general form of small velocity which corresponds to motion
of membranes. In Sec.3, we compute the one loop effective potential
around backgrounds with several velocities and discuss what BL theory
can probe about the target space. In Sec.4, we summarize and conclude. \\\\

During the preparation of present paper, we received the paper~\cite{Baek:2008ws}. The authors calculate 1-loop effective potential of ABJM theory and see an agreement with the Newton potential on $AdS_4 \times S^7/{\bf Z}_k$.

\vspace{10mm}

\section{Moduli space and position of Membranes}

In this note, we treat BL theory with $SO(4)$ gauge symmetry.
The moduli space of this theory has been studied in~\cite{Lambert:2008et, Distler:2008mk} at which the scalar potential vanishes. The Lagrangian is
\begin{align}
 {\cal L} &= -\frac{1}{2} D_\mu X^{A,I} D^\mu X_{A,I}
 +\frac{i}{2}\overline{\Psi}^A \Gamma^\mu D_\mu \Psi_A
 +\frac{ig}{4}f_{ABCD}\overline{\Psi}^B \Gamma^{IJ} X^{C,I}X^{D,J} \Psi^A
 \nonumber\\
 &\;\;\;
 -\frac{g^2}{12}\left[ f_{ABCD}X^{A,I}X^{B,J}X^{C,K}
 \right] \left[ f_{EFG}^{~~~~~D}X^{E,I}X^{F,J}X^{G,K}
 \right]
 \nonumber\\
 &\;\;\;
 +
  \frac{g}{2}\epsilon^{\mu\nu\lambda}\left[
  f_{ABCD}A_\mu^{AB}\partial_\nu A_{\lambda}^{CD}
  +\frac{2g}{3}f_{AEF}^{~~~~~G}f_{BCDG}A_\mu^{AB}A_\nu^{CD}A_\lambda^{EF}
 \right] ,
 \\
 D_\mu X^{A,I} &= \partial_\mu X^{A,I} + g \tilde{A}_{\mu B}^{A}X^{B,I}
 ,
 \hspace{3ex}
 \tilde{A}_{\mu B}^{A} \equiv f^A_{~~BCD}A_\mu^{CD}X^{B,I}
 ,
\end{align}
where $f_{ABCD}$ is the structure constant for the three algebra and
$f_{ABCD}=\epsilon_{ABCD}$, ($A=1,\cdots, 4$ etc), for ${\cal A}_4$ algebra which is equivalent with $SO(4)$ gauge symmetry. The indices $I,J,K (=1,\cdots, 8)$ are those of $SO(8)$ global symmetry and the spacetime signature is $(-,+,+)$. This Lagrangian has $N=8$ superconformal symmetry and supersymmetry requires the coupling constants are same and the value of coupling constant $g=2\pi/k$ is quantized ($k\in Z$), because of Chern-Simons term.

After a suitable gauge transformation, the vacuum configuration (with the gauge fields and fermions are zero) is 
\begin{align}
  \langle X^{A,I} \rangle &=\left(\begin{array}{c}
  0\\ 0\\ r_1^I \\ r_2^I  
  \end{array}\right)
  ,
  \hspace{3ex}
  \begin{array}{l}
  A=1,\cdots,4,
  \\
  I = 1,\cdots, 8,
  \end{array}
\end{align}
where $r_1^{I}$ and $r_2^{I}$ are real values, and the index $A=1,\cdots, 4$ is the index for the three algebra. 
There are two sets of eight values $r_1^I$ and $r_2^I$, and then $r_1^I$ and $r_2^I$ are related with the position of two membranes in the eight dimensional transverse directions in the target space.
The moduli space should be divided by the gauge symmetry. The discrete symmetry ${\cal O}(2,Z)\in SO(4)$ act on two vectors like:
\begin{align}
  \left(\begin{array}{cc}
  -1&0\\
  0 &1
  \end{array}\right) 
  &\hspace{3ex}
  :\hspace{3ex}
  r_1^I \rightarrow - r_1^I ,
  \hspace{3ex}
   r_2^I \rightarrow r_2^I ,
  \label{d1}
  \\
  \left(\begin{array}{cc}
  1&0\\
  0 &-1
  \end{array}\right) 
  &\hspace{3ex}
  :\hspace{3ex}
  r_1^I \rightarrow r_1^I , \hspace{3ex}
  r_2^I \rightarrow -r_2^I ,
  \\
  \left(\begin{array}{cc}
  ~0&1~\\
  ~1&0~
  \end{array}\right) 
  &\hspace{3ex}
  :\hspace{3ex}
  r_1^I \rightarrow r_2^I , \hspace{3ex}
  r_2^I \rightarrow r_1^I ,
  \label{d3}
\end{align}
and the moduli space becomes $( (R^8/Z_2)\times (R^8/Z_2) )/Z_2$. The moduli space should be further divided by the continuous gauge symmetry. Since the gauge fields have the Chern-Simons coupling, the continuous symmetry which keeps the Chern-Simons term invariant and $\tilde{A}_\mu^{A,B}=0$ is $Z_k\in U(1)$:
\begin{align}
  z^{I} \rightarrow e^{i\theta}z^I ,
  \hspace{3ex}
  \theta = \frac{\pi n}{k} , 
  \hspace{3ex}
  n\in Z ,
  \label{s1}
\end{align}
where $z^I = r_1^I + i r_2^I$. Then the moduli space is $( R^8\times R^8 )/D_{2k}$ where $D_{2k}$ is a dihedral group and for $k=1$ it is just $( R^8\times R^8 )/ ( Z_2\times Z_2 )$ and the target space is expected to $R^{1,2}\times (R^8/Z_2)$.
This $Z_k$ is a subgroup of $U(1)$, and this $U(1)$ generate one spacial direction and we call this direction the compactified direction even for a finite $k$. In the large $k$ limit, this direction is identified and the BL theory reduces to the weakly coupled IIA theory~\cite{Distler:2008mk}.

Using $SO(8)$ global symmetry, the form of $\langle X^{A,I} \rangle$ can be written
\begin{align}
  \langle X^{A,I} \rangle &=\left(\begin{array}{ccccc}
  0 &\cdots &0 & 0 & 0\\  
  0 &\cdots &0 & 0 & 0\\
  0 &\cdots &0 & b_0 & 0\\ 
  0 &\cdots &0 & 0 & a_0
  \end{array}\right)
  ,
\end{align}
\begin{figure}[t]
\begin{center}
\includegraphics[height=7cm]{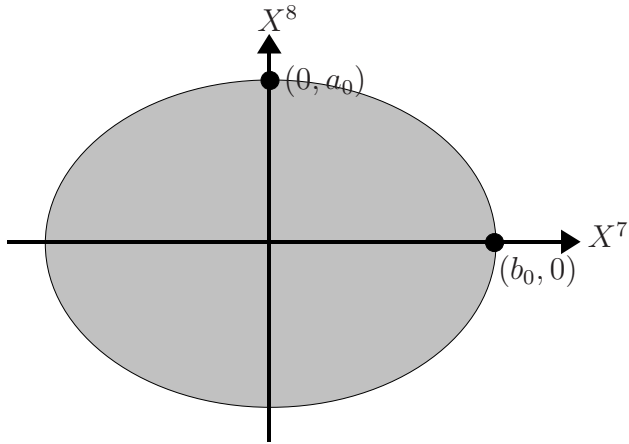}
\put(-26,104){$X^7$}
\put(-151,186){$X^8$}
\put(-60,92){$(b_0,0)$}
\put(-141,164){$(0,a_0)$}
\vspace{-5ex}
\caption{The positions of two membranes in $X^{I}$ coordinate, ($b_0,0$) and ($0,a_0$). The ellipse is the compactified direction generated by the $U(1)$ subgroup. The aria of the ellipse (the shaded region) is $\pi a_0b_0$.
}
\label{f1}
\end{center}
\end{figure}
and the position of membranes in $X^I$ coordinates and the compactified direction are plotted in Fig.~\ref{f1}.
 
When $a_0\neq 0$ and $b_0=0$, one can integrate out massive gauge fields and obtain $SU(2)$ (plus free $U(1)$) SYM theory, i.e. D2-brane action, at the leading order in $1/a_0$~\cite{Mukhi:2008ux}. If $b_0$ then turns on, $SU(2)$ gauge symmetry is broken down to $U(1)$ and the masses of massive gauge bosons are given $ga_0b_0$. Therefore $ga_0b_0=g_{YM}L$ where $g_{YM}$ is the gauge coupling of SYM and $L$ is the distance between two branes. Since only the product $g_{YM}L$ appears in the Lagrangian, there is an ambiguity for $g_{YM}$ (and $L$). We know there is a symmetry under the exchange of $a_0$ and $b_0$, and in $g\rightarrow 0, (k\rightarrow \infty)$ limit the theory reduces to the D2-brane system, and thus we choose $g_{YM}=g$ and $L=a_0b_0$ in this note.

Since we would like to discuss the scattering of membranes, we introduce the small time dependence into $X^{A,I}$. Solving the equations of motion for $\tilde{A}_\mu^{AB}$ and $X^{A,I}$ under $\tilde{A}_\mu^{A,B}=0$, we obtain
\begin{align}
  \langle X^{A,I} \rangle &=\left(\begin{array}{ccccc}
  0 &\cdots &0 & 0 & 0\\  
  0 &\cdots &0 & 0 & 0\\
  v_1t &\cdots &v_6t & b_0+v_7t & v_8t\\ 
  u_1t &\cdots &u_6t & u_7t & a_0+u_8t
  \end{array}\right)
  ,
\end{align}
and except that $v_8=(b_0u_7)/a_0$, all the $v$ and $u$ are free. We note that the constraint $v_8=(b_0u_7)/a_0$ comes from the equation of motion for $\tilde{A}^{3,4}_\mu$ which is the gauge field corresponds to the continuous symmetry~\eqref{s1} and means that the momentum along the compactified direction is set to be zero. This may be the similar situation to that one light-cone direction is compactified and the momentum along that direction is set to be constant in BFSS matrix model.

\section{Membranes scattering and gravitational potential}

In the previous section, we review the moduli space and the general form of velocity which satisfies the equation of motion. In this section we study the one loop effective potential around the background with non-zero velocities. In string theory, D-brane scattering has been discussed using SYM theory and the one loop effective potential reproduces the Newton potential in the target space. Thus we expect we can probe the target space from the one loop potential which we will compute in this section.

From the relation which comes from the gauge field $\tilde{A}_\mu^{3,4}$, one spacial direction is special and there is no momentum transfer along the direction. Although we expect that the target space is eleven dimensions, this observation implies we can only probe ten dimensions, not eleven dimensions. On the other hand, the action has $SO(8)$ global symmetry and (supersymmetric) conformal symmetry and we may expect we can probe eleven dimension according to the discussion by~\cite{Banks:1997uu}. Thus we compute the gravitational potential by applying the idea of computing the gravitational potential from SYM theory, to clarify which observation is correct.\\

Before going to the calculation,
we notice that the regularization in Chern-Simons theory is not simple. A dimensional regularization naively breaks the gauge invariance due to the difficulty of analytic continuation of $\epsilon_{\mu\nu\rho}$. Another regularization is adding Yang-Mills term and a careful study on the regularization methods has been done in~\cite{Chen:1992ee}. The one loop corrections in BL theory have been discussed  with these regularization procedure~\cite{Gustavsson:2008bf}. In our calculation of one loop graphs, a dimensional regularization can be adapted.

\subsection{For $v_7\neq 0$ and $u_8\neq 0$}

We first study the case where only $v_7$ and $u_8$ are non-zero. In
order to study the one loop effective potential, we just have to keep
quadratic terms in the Lagrangian around the background. Then the relevant terms in the Lagrangian becomes $ {\cal L} ={\cal L}_1+{\cal L}_2 +{\cal L}_f$,
\begin{align}
 {\cal L}_{1}&=
 \sum_{\alpha=1,2} g \epsilon^{\mu\nu\rho} \tilde{A}^{\alpha}_\mu \partial_\nu \tilde{B}^{\alpha}_\rho
 -\frac{1}{2} [ \partial_\mu X^{2,7} + g b \tilde{A}^1_{\mu} ]^2
 -(\partial_t b) g \tilde{A}^1_{t} X^{2,7} 
 -\frac{1}{2} [ \partial_\mu X^{1,8} + g a \tilde{B}^1_{\mu} ]^2
 -( \partial_t a) g \tilde{B}^1_{t} X^{1,8}
\nonumber\\
 &\;\;\;
 -\frac{1}{2} [ \partial_\mu X^{1,7} - g b \tilde{A}^2_{\mu} ]^2
 +(\partial_t b) g \tilde{A}^2_{t} X^{1,7}
 -\frac{1}{2} [ \partial_\mu X^{2,8} + g a \tilde{B}^2_{\mu} ]^2
 -( \partial_t a) g \tilde{B}^2_{t} X^{2,8}
 ,
 \nonumber
 \\
 {\cal L}_2&=
 g \epsilon^{\mu\nu\rho} \tilde{A}^3_\mu \partial_\nu \tilde{B}^3_\rho
 -\frac{1}{2} [ \partial_\mu X^{3,7} ]^2
 -\frac{1}{2} [ \partial_\mu X^{3,8} + g a \tilde{B}^3_{\mu} ]^2
 -( \partial_t a) g \tilde{B}^3_{t} X^{3,8} 
 -\frac{1}{2} [ \partial_\mu X^{4,7} - g b \tilde{B}^3_{\mu} ]^2
 \nonumber
 \\
 &\;\;\;
 +(\partial_t b) g \tilde{B}^3_{t} X^{4,7} 
 -\frac{1}{2} [ \partial_\mu X^{4,8} ]^2
 ,
\nonumber
\\
 {\cal L}_f &=
 \frac{1}{2}
 X^{A,I} ( \Box -g^2a^2b^2 ) X^{A,I}
 +\frac{1}{2}
 X^{A',I} \Box X^{A',I}
 +\overline{\Psi}^{A''} \Gamma^\mu \partial_\mu \Psi_{A''}
 +\frac{i}{2}gab [ \overline{\Psi}^2 \Gamma^{78} \Psi^1 - \overline{\Psi}^1 \Gamma^{78} \Psi^2 ]
  ,
 \nonumber
 \\
 &\;\;\;
  (A=1,2,
  \hspace{2ex} 
  A'=3,4,
    \hspace{2ex} 
  A''=1,\cdots,4,
    \hspace{2ex} 
  I=1,\cdots,6 
)
  \nonumber
\end{align}
where $a\equiv a_0+u_8t$ and $b\equiv b_0+v_7t$, and we have used the following notation
\begin{align}
 \tilde{A}_\mu^\alpha \equiv \frac{1}{2}\epsilon^{\alpha\beta\gamma}\tilde{A}_{\mu \beta\gamma}
 ,\hspace{3ex}
 \tilde{B}_\mu^\alpha \equiv \tilde{A}_\mu^{\alpha 4}
 ,
 \hspace{3ex}
 ( \alpha = 1,\cdots,3, \mbox{ etc} )
 .
\end{align}
\begin{figure}[t]
\begin{center}
\includegraphics[height=7cm]{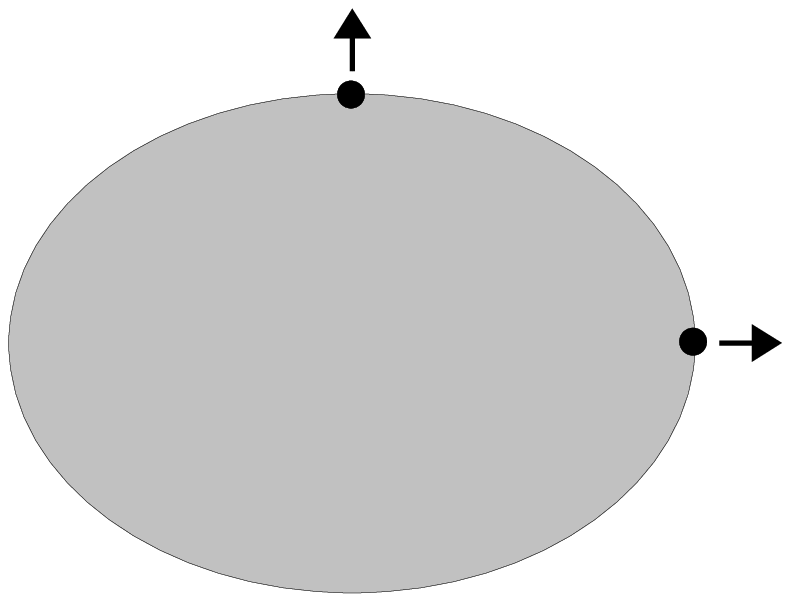}
\vspace{-7ex}
\caption{The arrows denote the direction of velocity.
}
\label{f2}
\end{center}
\end{figure}
\noindent
In this case the direction of velocity is normal to the compactified direction (Fig.~\ref{f2}).

We integrate out $\tilde{A}_\mu^3$ which gives that $\tilde{B}_\mu^3$ is written by a derivative of scalar field, i.e. $\tilde{B}_\mu^3=\partial_\mu B$. 
Substitute this expression into ${\cal L}_2$, we obtain
\begin{align}
 {\cal L}_2&=
 -\frac{1}{2} [ \partial_\mu X^{3,7} ]^2
 -\frac{1}{2} [ \partial_\mu X^{3,8} + g (\partial_\mu aB) ]^2
 -\frac{1}{2} [ \partial_\mu X^{4,7} - g (\partial_\mu bB) ]^2
 -\frac{1}{2} [ \partial_\mu X^{4,8} ]^2
 ,
\end{align}
after by using a partial integral. Thus we have four massless scalar fields and the contribution to the one loop effective action from this part becomes
\begin{align}
 V_2^{\rm 1\; loop}(a_0,b_0;u_8,v_7) = \int \! d^3x \; 4\times \frac{1}{2} \ln \det \Box
 .
\end{align}
We can also easily compute the contribution from ${\cal L}_f$ which are twelve massive scalars with the mass $gab$, sixteen massless fermion and eight massive fermions with the mass$^2$ $g^2a^2b^2\pm g\partial_t (ab)$. Then we obtain 
\begin{align}
 V_f^{\rm 1 \; loop}(a_0,b_0;u_8,v_7) &=  \int \! d^3x \; 
 12\times \frac{1}{2} \ln \det ( \Box - g^2a^2b^2 )
 +12\times \frac{1}{2} \ln \det \Box
 -16\times \frac{1}{2} \ln \det \Box
 \nonumber \\
 &\;\;\;\; \;\;\;\; 
 -8\times \frac{1}{2} \Big[ \ln \det ( \Box - g^2a^2b^2 +g(\partial_t ab) ) 
 + \ln \det ( \Box - g^2a^2b^2 -g(\partial_t ab) ) \Big]
 .
\end{align}
Now we study ${\cal L}_{1}$. We similarly integrate out $\tilde{B}_\mu^{a}$ using the equation of motion and we obtain,
\begin{align}
 {\cal L}_{1}&= -\frac{1}{4a^2} [ \partial_{\mu}\tilde{A}_{\nu}^{1} - \partial_{\nu}\tilde{A}^{1}_{\mu} ]^2 - \frac{1}{2}[\partial_{\mu} X^{2,7}+gb\tilde{A}_{t}^{1}]^2
 +(\partial_t b)g\tilde{A}^{1}_{t}X^{2,7}
 \nonumber \\ 
 &\;\;\;
 -\frac{1}{4a^2} [ \partial_{\mu}\tilde{A}_{\nu}^{2} - \partial_{\nu}\tilde{A}_{\mu}^{2} ]^2-\frac{1}{2}[\partial_{\mu} X^{1,7}-gb\tilde{A}_{2}]^2
 -(\partial_t b)g\tilde{A}_{t}X^{1,7}
 .
\end{align}
The Lagrangian ${\cal L}_1$ is exactly same as the quadratic part of two D2-brane action with the time dependent gauge coupling $a$.  Thus we immediately see that if $u_8=0$, the one loop effective potential is exactly same as that of two D2-brane scattering with the gauge coupling $a_0$ and the distance 
between two D2 brane in $X^7$ direction is $gb$. Then in this case, we have
\begin{align}
 V_{1}^{\rm 1 \; loop}(a_0,b_0;u_8=0,v_7) &= \int \! d^3x \; 
 2\times \frac{1}{2} \Big[
 \ln \det ( \Box - g^2a^2b^2 +(2g\partial_t ab) )
 \nonumber \\
 &\;\;\;\; \;\;\;\;
 +\ln \det ( \Box - g^2a^2b^2 -(2g\partial_t ab) )
 \Big]
 ,
\end{align}
and in total the one loop effective potential by expanding $v_7t \ll b_0$ is
\begin{align}
 V^{\rm 1 \; loop}(a_0,b_0;u_8=0,v_7) &= \int \! \frac{d^3p}{(2\pi)^3} \;
 \frac{ 2 (ga_0\partial_t b)^4 }{ ( p^2+g^2a_0^2b^2 )^4}
 +{\cal O}((v_7t)^6). 
\end{align}
We notice that the terms with the second order in $v_7$ cancel out. 
It gives the following potential at the leading order 
\begin{align}
 V^{\rm 1 \; loop}(a_0,b_0;u_8=0,v_7) &=
 c_{YM} \frac{ (a_0\partial_t b)^4 }{ ga_0^5b_0^5 }
 = c_{YM} \frac{ (\partial_t L)^4 }{ g_{YM}L^5 },
\end{align}
where $c_{YM}$ is the numerical coefficient computed from D2-brane scattering using SYM theory, and $g_{YM}=g$ and $L=a_0b$ from the matching with D2-brane action in the $g\rightarrow 0$ limit.
Since there is a discrete symmetry which exchange $a$ and $b$, the one loop effective potential of the case $v_7=0$ is same as that of two D2-brane scattering with the gauge coupling $b_0$ and the distance $a$. Then in this case we have a same form
\begin{align}
 V^{\rm 1 \; loop}(a_0,b_0;u_8,v_7=0) &= \int \! \frac{d^3p}{(2\pi)^3} \;
 \frac{ (g\partial_t ab_0)^4 }{ ( p^2+g^2a^2b_0^2 )^4}
 \sim c_{YM}\frac{ (\partial_t ab_0)^4 }{ ga_0^5b_0^5 }
 = c_{YM}\frac{ (\partial_t L)^4 }{ g_{YM}L^5}
 ,
\end{align}
where $g_{YM}=g$ and $L=ab_0$.

For both $v_7$ and $u_8$ are non-zero, the calculation is involved and we introduce a proper gauge fixing term and compute the one loop effective potential. In order that the computation becomes simple, first we rescale $\tilde{A}_{\mu}^{\alpha=1,2}= aA_{\mu}^{\alpha=1,2}$ to have canonical kinetic terms
 $\frac{1}{2}A^{\alpha}[\Box-(gab)^2]A^{\alpha}$.
Next, we introduce the following gauge fixing
\begin{align}
 {\cal L}_{gf} &=
-\frac{1}{2}\left[\partial^{\mu}A^{1}_{\mu}+gabX^{2,7}-\frac{ \partial^{\mu}a}{a} A^{1}_{\mu}\right]^2
-\frac{1}{2}\left[\partial^{\mu}A^{2}_{\mu}-gabX^{1,7}-\frac{\partial^{\mu}a   }{a^2}A^{2}_{\mu}\right]^2.
 \end{align}
 The ghost Lagrangian may be suggested from ${\cal L}_{gf}$ as
\begin{eqnarray}
{\cal L}_{gh} = \sum_{\alpha=1,2}
\tilde{\bar{c}}^{\alpha}a\left[\partial^{\mu} \frac{1}{a}\partial_{\mu} -g^2ab^2-\frac{\partial^{\mu} a}{a^2}\partial_{\mu} \right]\tilde{c}^{\alpha}
,
\label{ghlag}
\end{eqnarray}
which follows from the gauge symmetry of ${\cal L}_{1}$:
\begin{eqnarray}
&&\delta A^{1}_{\mu}=\frac{1}{a}\partial_{\mu}\Lambda^1, \quad \delta X^{2,7}=-gb\Lambda^1. 
\nonumber \\
&&\delta A^{2}_{\mu}=\frac{1}{a}\partial_{\mu}\Lambda^2, \quad \delta X^{2,8}=gb\Lambda^2.
\end{eqnarray} 
However, note that these ghost may allow a background dependent field rescaling
$\tilde{c}\rightarrow f(a,b)\tilde{c}$ and $\tilde{\bar{c}}\rightarrow f(a,b)^{-1}\tilde{\bar{c}}$ with some function $f(a,b)$. The normalization is fixed such that the ghost Lagrangian has the supersymmetry after adding superpartners appropriately. In stead of fixing the normalization from supersymmetry, we can determine the correct normalization from the requirement that the total Lagrangian has the discrete symmetry under the exchange~\eqref{d1}-\eqref{d3}. It 
is simply achieved by the ghost redefinition $\tilde{c}=ac$ and $\tilde{\bar{c}}=a^{-1}\bar{c}$ in (\ref{ghlag}), then we claim that correct ghost Lagrangian is
\begin{eqnarray}
{\cal L}_{gh}=\sum_{\alpha=1,2}\bar{c}^{\alpha}\left[\Box-g^2a^2b^2-2\frac{\partial^{\mu}a\partial_{\mu}a}{a^2}\right]c^{\alpha}.
\label{aght}
\end{eqnarray}
Then ${\cal L}_1+{\cal L}_{gf}+{\cal L}_{gh}$ becomes 
\begin{eqnarray} 
&&{\cal L}_1+{\cal L}_{gf}+{\cal L}_{gh} =  
\sum_{\alpha=1,2}\frac{1}{2}X^{\alpha,7}(\Box-g^2a^2b^2)X^{\alpha,7}  +2g\partial^{\mu}(ab)\tilde{A}^{1}_{\mu}X^{2,7}
-2g\partial^{\mu}(ab)\tilde{A}^{2}_{\mu}X^{1,7}
\nonumber \\
&&\hspace{6ex}
 +\sum_{\alpha=1,2}\left[
\frac{1}{2}\tilde{A}^{\alpha\;\mu}(\Box-g^2a^2b^2)\tilde{A}^{\alpha}_{\mu}+\frac{1}{2}\frac{\partial^{\mu}a\partial^{\nu}a}{a^2}\tilde{A}^{\alpha}_{\mu}\tilde{A}^{\alpha}_{\nu}-\frac{\partial^{\mu}a\partial_{\mu}a}{a^2}\tilde{A}^{\alpha\;\mu}\tilde{A}^{\alpha}_{\mu}\right] 
\nonumber \\
&&\hspace{6ex}
 +\sum_{\alpha=1,2}\bar{c}^{\alpha}\left[\Box-g^2a^2b^2-2\frac{\partial^{\mu}a\partial_{\mu}a}{a^2}\right]c^{\alpha}
 .
\end{eqnarray} 
We compute the one loop effective action as  a perturbation with  $v=\partial_t(ab)$ and $\partial_t a$. The terms proportional to $(\partial_t a)^2$ and $(\partial_t a)^4$ cancel out between the gauge fields and ghosts, and because of this the ghost action~\eqref{aght} is consistent with the discrete symmetry~\eqref{d1}-\eqref{d3}. We can see the second order in terms of velocity $v=\partial_t(ab)$ cancels out as expected from supersymmetry. 
This is because the boson loop contribution from ${\cal L}_{1}+{\cal L}_{gf}+{\cal L}_{gh}+{\cal L}_f$ is
\begin{align}
\int \! \frac{d^3p}{(2\pi)^3} \; \frac{4g^2[\partial(ab)]^2}{(p^2+g^2a^2b^2)^2}
\end{align}
and it is canceled by the fermion loop contribution from ${\cal L}_f$.
We can also easily see that the third order of $v$ vanishes and the potential starts from the fourth order in $v$,
\begin{align}
 V^{\rm 1\;loop}(a_0,b_0;u_8,v_7) &\sim c_{YM} \frac{[(ub_0+a_0v)]^4}{ga_0^5b_0^5}
 = c_{YM}\frac{(\partial_t L)^4}{g_{YM}L^5}
 .
\end{align}
In summary, we obtain that the form of one loop effective potential is given by $(\partial_t L)^4/L^5$  and the exponent $5$ for $L$ is consistent with the gravitational potential in ten dimensional space.

\vspace{2ex}

From the above result that there are no $1/a^6$ or $1/b^6$ terms in the potential, when $a_0\neq 0$ and $b_0=0$ the potential vanishes at the one loop,
\begin{align}
 V^{\rm 1\;loop}(a_0,b_0=0;u_8,v_7=0) &= 0
 . \label{a-pote}
\end{align}
Since Bagger-Lambert theory is a superconformal theory, the canonical dimension of $a$ is half and the possible form for the potential has the following form
\begin{align}
 V_{\rm eff}(a_0,b_0=0;u_8,v_7=0) \propto \frac{\partial_t a}{a^6}
 , 
\end{align}
If the coefficient is not zero, we may claim that $a$ is the distance between two membranes and the target space is a eleven dimensional space from a similar argument on a scale invariant $SO(8)$ symmetric theory~\cite{Banks:1997uu}. But, as (\ref{a-pote}), the coefficient is zero in Bagger-Lambert theory. 

\vspace{1ex}

These results suggest that the membrane fluctuations  connecting two
membranes always wrap the compactified direction generated by $\tilde{A}_\mu^{3,4}$  even $k$ is finite, and therefore the one loop effective potential only probes ten dimensions. Since in the large $a_0$ limit the action at the leading terms in $1/a_0$ is same as the action for D2-branes, this result is natural.

\subsection{For $v_8\neq 0$ and $u_7\neq 0$}

\begin{figure}[t]
\begin{center}
\includegraphics[height=7cm]{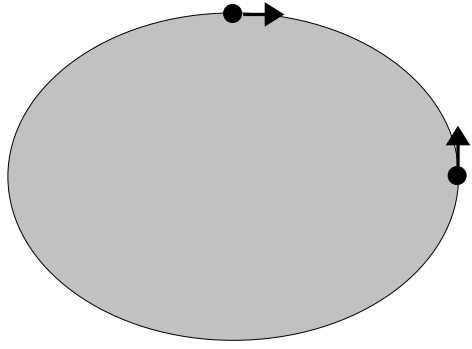}
\vspace{-7ex}
\caption{The arrows denote the direction of velocity.
}
\label{f3}
\end{center}
\end{figure}

We study the case $v_8\neq 0$ and $u_7=a_0v_8/b_0\neq 0$. In the previous case,  the membranes are pulled normal to the compactified direction. On the other in the case $v_8\neq 0$ and $u_7\neq 0$, the direction of velocity is tangent to the compactified direction (Fig.~\ref{f3}), (but notice that the momentum along the compactified direction is always zero).
\noindent
We may expect the result is different from the previous case.

The relevant term of Lagrangian after the redefinition 
$\tilde{A}_\mu^\alpha = a_0A_\mu^\alpha$ and $\tilde{B}_\mu^\alpha = b_0 B_\mu^\alpha$ becomes $ {\cal L} = {\cal L}_1 + {\cal L}_2 + {\cal L}_f$,
\begin{align}
 {\cal L}_1 &=
  \sum_{\alpha=1,2} gL \epsilon^{\mu\nu\rho} A^{\alpha}_\mu \partial_\nu B^{\alpha}_\rho
 -\frac{1}{2} [ \partial_\mu X^{1,7} - gL {A}^2_{\mu} + gVt{B}_\mu^1]^2
 -\frac{1}{2} [ \partial_\mu X^{2,7} + gL {A}^1_{\mu} + gVt{B}_\mu^2 ]^2
 \nonumber \\
 &\;\;\;
 -\frac{1}{2} [ \partial_\mu X^{1,8} + gL {B}^1_{\mu}  - gVt{A}_\mu^2]^2
 -\frac{1}{2} [ \partial_\mu X^{2,8} + gL {B}^2_{\mu}  + gVt{A}_\mu^1]^2
 \nonumber \\
 &\;\;\;
 - V g ( {B}_t^1 X^{1,7} + {B}_t^2 X^{2,7}) 
 - V g ( -{A}_t^2 X^{1,8} + {A}_t^1 X^{2,8})
 ,
 \label{L1}
 \\
 {\cal L}_2 &=
 gL \epsilon^{\mu\nu\rho} {A}^3_\mu \partial_\nu {B}^3_\rho
 - V g {B}_t^3 X^{3,7} + V g {B}_t^3 X^{4,8}
 -\frac{1}{2} [ \partial_\mu X^{3,7} + gVt{B}_\mu^3]^2
 -\frac{1}{2} [ \partial_\mu X^{4,8} - gVt{B}_\mu^3]^2
 \nonumber \\
 &\;\;\;
 -\frac{1}{2} [ \partial_\mu X^{4,7} - gL {B}^3_{\mu} ]^2
 -\frac{1}{2} [ \partial_\mu X^{3,8} + gL {B}^3_{\mu} ]^2
 ,
 \label{L3}
 \\
 {\cal L}_f &=
 \frac{1}{2} X^{A,I} \Big[ \Box -g^2 (L - \frac{V^2}{L}t^2 )^2 \Big] X^{A,I}
 +\frac{1}{2}  X^{A',I} \Box X^{A,I}
 +\frac{i}{2} \overline{\Psi}^{A''} \Gamma^\mu \partial_\mu \Psi_{A''}
 \nonumber
 \\
 &\;\;\;
 +\frac{i}{2}g(L-\frac{V^2}{L}t^2) ( \overline{\Psi}^2 \Gamma^{78} \Psi^1 - \overline{\Psi}^1 \Gamma^{78} \Psi^2 )
 ,
 \nonumber
 \\
 &\;\;\;
 (A=1,2, \hspace{2ex}
 A'=3,4, \hspace{2ex}
 A''=1,\cdots,4 \hspace{2ex}
 I=1,\cdots,6 ),
 \label{Lf}
\end{align}
where $L=a_0b_0$ and $V=a_0v_8=b_0u_7$. The background always appears in the combination $L$ and $V$ and this Lagrangian can not be understood as SYM with time dependent gauge coupling and/or time dependent Higgs fields after integrating out $B_\mu^\alpha$ fields. 
In this case the 1-loop effective potential becomes (We discuss on the calculation of the potential in appendix~A.)
\begin{eqnarray}
 V^{\rm 1\;loop}(L,V)&=&\frac{1}{g\pi}\int d^3x\;\left[\frac{V^4}{4L^5}-\frac{g^2V^4}{2L^3}t^2\right].
\end{eqnarray}
In $u_7\rightarrow 0$ with fixed $V$ limit it becomes D2-brane like potential. 
Again it is suggested that 2-branes feel large ten dimension through this potential. \\

\vspace{2ex}

Before closing this section, we give two comments.

(1) Since the physical mass scale is $ab-u_7v_8t^2$, we expect that if $ab-u_7v_8t^2=a_0b_0$ is kept fixed the effective
potential is zero. However $ab-u_7v_8t^2=a_0b_0$ implies all the velocities $u_{7,8}$ and $v_{7,8}$ are zero.

(2) We look again at the equations of motion for the gauge fields $\tilde{B}_\mu^\alpha$ and $\tilde{A}_\mu^\alpha$ (with fermions are zero) which are
\begin{align}
  0 & =
   - X^{4,I} D_\mu X^{\alpha,I} 
  + X^{\alpha,I} D_\mu X^{4,I} 
  +\frac{1}{2} \epsilon^{\mu\nu\rho} ( \tilde{F}_{\nu\rho}^\alpha -g\epsilon^{\alpha\beta\gamma}\tilde{B}_\nu^\beta\tilde{B}_\rho^\gamma   ) ,
  \label{be}
  \\
  0 & =
   \epsilon_{\alpha\beta\gamma} X^{\beta,I} D_\mu X^{\gamma,I} 
  +\epsilon^{\mu\nu\rho} ( \partial_\nu \tilde{B}_\rho^\alpha -g\epsilon^{\alpha\beta\gamma}\tilde{A}_\nu^\beta\tilde{B}_\rho^\gamma   ) ,
\end{align}
where $\tilde{F}^{\alpha}_{\mu\nu}$ is $SU(2)$ gauge field strength constructed by $\tilde{A}_\mu^\alpha$. 
Then the momentum along the compactified direction is non zero
($-X^{4,I}D_\mu X^{3,I}+X^{3,I}D_\mu X^{4,I}\neq 0$)
 when $\tilde{A}_\mu^\alpha$ has a magnetic monopole configuration (with $\tilde{B}_\mu^\alpha =0$). (The monopole instanton configuration in ABJM theory is discussed in~\cite{Hosomichi:2008ip}.) This is consistent with the membrane scattering from three dimensional SYM with M-momentum discussed by Polchinski-Pouliot~\cite{Polchinski:1997pz}. Therefore we expect that higher loop contributions do not change the form of leading potential, $\propto (\partial_t L)^4/L^5$ and the eleventh direction cannot be probed perturbatically. The eleventh direction can be probed through a non-perturbative process.

\section{Conclusion and Discussions}

In this note, we studied membrane scattering from Bagger-Lambert theory and read out the dimensions of the target space from the one loop effective potential. We understand the membranes propagating between two membranes always wrap on the one spacial direction which becomes the compactified direction when the level of Chern-Simons coupling $k$ becomes infinite. This special direction cannot be probed and the membrane can only probe ten dimensions in perturbation, though the Bagger-Lambert theory has $SO(8)$ and scale symmetries. As similar to the membrane scattering from SYM theory, the eleventh direction can be probed through non perturbative effects.

\bigskip\bigskip

\subsection*{Acknowledgements}

The authors would like to thank all the members of string group in Taiwan and especially Pei-Ming Ho for useful discussion. This work is supported by National Center for Theoretical Sciences, Taiwan,
(No. NSC 97-2119-M-002-001, NSC97-2119-M-007-001).

\bigskip\bigskip

\appendix

\section{Detail of the one loop potential in $v_8\neq 0, u_7\neq 0$ case }

Here we discuss the effective potential in section 3.2 in some detail. 
Contributions from ${\cal L}_f$ in (\ref{Lf}) are simple. 
Contributions from ${\cal L}_2$ in (\ref{L3}) can be written as those of four massless scalars after integrating out $B^3_{\mu}$, as similar to section 3.1. So let us consider ${\cal L}_1$ in  (\ref{L1}) in bellow. 

To make the calculation  easy first we integrate out $B^\alpha_{\mu}$ in (\ref{L1}) and redefine scalar field as
\begin{eqnarray}
\left(\begin{array}{c}
X^{\alpha,7} \\X^{\alpha,8}
\end{array}\right)=
\left(\begin{array}{cc}
L&Vt \\
-Vt&L
\end{array}\right)
\left(\begin{array}{c}
X^\alpha \\Y^\alpha
\end{array}\right), \qquad \alpha=1,2.
\end{eqnarray}
Then  (\ref{L1}) becomes
\begin{align}
 {\cal L}_1 &=
 -\frac{1}{2}(L^2+V^2t^2)(\partial_\mu X^{\alpha})^2
 +2\frac{[\partial(Vt)]^2}{\xi^2}(X^\alpha)^2
  -\frac{g^2\xi^2}{2}(L^2-V^2t^2)(X^\alpha)^2
 -\frac{1}{2\xi^2}\partial_{\mu}A_{\nu}^{\alpha}\partial^{\mu}A^{\alpha\;\nu}
 \nonumber \\
 &\;\;\;\;
 +\Big[\frac{\partial^{\mu}\xi\partial^{\nu}\xi}{\xi^4}-\frac{\partial^{\mu}\partial^{\nu}\xi}{\xi^3}\Big] A^\alpha_{\mu}A^\alpha_{\nu}
 -\frac{1}{2}g^2\left[ L^2 \xi^2
-\frac{4V^2t^2}{\xi^2} \right](A^\alpha)^2 
  +\frac{4gVt}{\xi^2}{\epsilon_{\mu}}^{\nu\rho}\partial_{\nu}A^{1,2}_{\rho}A^{2,1}_{\mu}
 \nonumber \\
 &\;\;\;\;
+\frac{1}{2\xi^2}\left[\partial^{\mu}A^{1}_{\mu}-\frac{2\partial^{\mu}\xi}{\xi}A_{\mu} + g\xi^2(L^2-V^2t^2)X^{2}\right]^2 
+\frac{1}{2\xi^2}\left[\partial^{\mu}A^{2}_{\mu}-\frac{2\partial^{\mu}\xi}{\xi}A_{\mu} - g\xi^2(L^2-V^2t^2)X^{1}\right]^2 
 \nonumber \\
&\;\;\;\;
 +\frac{2\partial_{\mu}(Vt)}{\xi^2}\epsilon^{\mu\nu\rho}\partial_{\nu}A^{1,2}_{\rho}X^{1,2}
 \mp g\partial^{\mu}(V^2t^2) \Big( 1 + \frac{2}{\xi^2} \Big) A^{1,2}_{\mu}X^{2,1}
 \pm2g(L^2-V^2t^2)\frac{\partial^{\mu}\xi}{\xi}A_{\mu}^{1,2}X^{2,1}
 ,
\end{align}
where $\xi^2=1+\frac{(Vt)^2}{L^2}$. In the last line, we introduced a convenient notation: $A^{1,2}X^{1,2}=A^{1}X^{1}+A^{2}X^{2}$,   $\pm A^{1,2}X^{2,1}=A^{1}X^{2}-A^{2}X^{1}$.
Note that $Y^a$ disappeared from the Lagrangian due to a Higgs mechanism.  
Next we introduce a gauge fixing Lagrangian
\begin{eqnarray}
{\cal L}_{gf}&=&\frac{1}{2\xi^2}\left[\partial^{\mu}A^{1,2}_{\mu}
-\frac{2\partial^{\mu}\xi}{\xi}A_{\mu}^{1,2}
\pm g\xi^2(L^2-V^2t^2)X^{2,1}\right]^2.
\end{eqnarray} 
Then a naive ghost Lagrangian (we will explain later why this Lagrangian is naive) would be
\begin{eqnarray}
{\cal L}_{gh}= \sum_{\alpha=1,2}
 \bar{\tilde{c}}^{\alpha}\left[
\Box-\frac{2\partial^{\mu}\xi}{\xi}\partial_{\mu}
-g^2L^2\left(1-\frac{V^2t^2}{L^2}\right)^2
\right]\tilde{c}^{\alpha}.
\end{eqnarray} 
Finally we make kinetic terms of $X^\alpha$ and $A^\alpha_{\mu}$ 
canonical by a field rescaling, and then the gauge fixed Lagrangian is
\begin{eqnarray}
 && {\cal L}_{1}+{\cal L}_{gf}+{\cal L}_{gh}= 
 \nonumber \\ 
 &&
 -\frac{1}{2}(\partial_\mu X^{\alpha})^2
 -\frac{1}{2}\left[ g^2L^2\left(1-\frac{V^2t^2}{L^2}\right)^2
 +\frac{\Box\xi}{\xi}-\frac{4\partial^{\mu}(Vt)\partial_{\mu}(Vt)}{L^2}
 \right](X^\alpha)^2 
 \nonumber \\
&&-\frac{1}{2}(\partial_{\mu}A^\alpha_{\nu})(\partial^{\mu}A^{\alpha\;\nu})
+\left(\frac{\partial^{\mu}\xi\partial^{\nu}\xi}{\xi^2}-\frac{\partial^{\mu}\partial^{\nu}\xi}{\xi}\right)A^\alpha_{\mu}A^\alpha_{\nu}
-\frac{1}{2}\left[ g^2L^2\xi^4-4g^2V^2t^2 
-\frac{\Box\xi}{\xi}
\right](A^\alpha)^2 \nonumber \\
&&+\bar{\tilde{c}}^\alpha\left[
\Box-\frac{2\partial^{\mu}\xi}{\xi}\partial_{\mu}
-g^2L^2\left(1-\frac{V^2t^2}{L^2}\right)^2
\right]\tilde{c}^\alpha 
\pm\frac{4gV}{\xi^2} \epsilon^{\mu\nu\rho}\partial_{\nu}
(\xi A^{1,2}_{\rho})A^{2,1}_{\mu} 
\nonumber \\
&&
+\frac{2\partial_{\mu}(Vt)}{L\xi^2}\epsilon^{\mu\nu\rho}\partial_{\nu}(\xi
A^{1,2}_{\rho})X^{1,2} 
 \mp 2gL\frac{\partial^\mu (V^2t^2)}{L}\left(1+\frac{2}{\xi^2}\right) A^{1,2}_{\mu}X^{2,1}
\pm2g(L^2-V^2t^2)\frac{\partial^{\mu}\xi}{\xi L}A_{\mu}^{1,2}X^{2,1}
. \nonumber \\ \label{gLwm} 
\end{eqnarray}
Then we calculate 1-loop effective potential as a perturbation of $V$. 
Now let us calculate $O(V^2)$ terms of 1-loop potential by using this Lagrangian. Interaction vertices which are relevant for our calculation are
\begin{align}
 {\cal V} &=
X^\alpha\left[(gVt)^2-\frac{3V^2}{2L^2}\right]X^\alpha
+A^{\alpha\;\mu}\left[(gVt)^2-\frac{V^2}{2L^2}\right]A^\alpha_{\mu}
+\frac{V^2}{L^2}A_{0}^{\alpha}A_{0}^{\alpha}
 \nonumber \\
 &\;\;\;
 +\bar{c}^\alpha\left[2(gVt)^2+\frac{2V^2t}{L^2}\partial_t\right]c^\alpha 
 +2gV\epsilon^{ij}\epsilon A^{1}_{i}A^{2}_{j}
 -\frac{2V}{L}X^{1,2}\epsilon^{ij}\partial_{i}A^{1,2}_{j}, 
 \hspace{3ex} 
 (i,j=1,2) 
\end{align}
and free field propagators are
\begin{eqnarray}
\langle X^{\alpha}(x)X^{\beta}(y)\rangle=\delta^{\alpha,\beta}\Delta(x,y), \qquad 
\langle A^{\alpha}_{\mu}(x)A^{\beta}_{\nu}(y)\rangle=\delta^{\alpha,\beta}\eta_{\mu\nu}\Delta(x,y)
,
\end{eqnarray}
where 
\begin{eqnarray}
\Delta(x,y) =\int \frac{d^3p}{i(2\pi)^3}\frac{e^{ip(x-y)}}{p^2+(gL)^2},
\end{eqnarray}
which satisfies
\begin{eqnarray}
(\Box^{(x)}-g^2L^2)\Delta(x,y)=i\delta^{(3)}(x-y).
\end{eqnarray}
A simple calculation shows that $O(V^2)$ terms are
\begin{eqnarray}
 -4i\frac{V^2}{L^2}\int dx^3(1+t\partial_2)\Delta(x,x)
 +\frac{4V^2}{L^2}\int dx^3\int dy^3\;\Delta(x,y)(\partial^2_{x_i}-g^2L^2)\Delta(x,y).
\label{V2term}
\end{eqnarray} 
The contributions from fermionic loop cancels by themselves and the total potential is given by (\ref{V2term}).
This seems to contradict with supersymmetry since $V^2$ should vanish because of supersymmetry. This is because the ghost Lagrangian was naive.
Namely,  the normalization of ghost fields has not been fixed yet, and
one may determine the normalization so that the
result is consistent with supersymmetry. Rescaling ghost fields as $\tilde{c}=fc$ and $\bar{\tilde{c}}=f^{-1}\bar{c}$, we have a new derivative interaction 
\begin{eqnarray}
\delta{\cal V}=\bar{c}\left[
\frac{2(\partial^{\mu}f)\partial_{\mu}}{f}+\frac{(\Box f)}{f}
\right]c
,
\end{eqnarray}
to ghost Lagrangian ${\cal L}_{gh}$. And here we propose to choose
\begin{eqnarray}
 f = 1 \pm i \frac{Vt}{L}
 ,
\end{eqnarray}
and this interaction terms cancels $O(V^2)$ term (\ref{V2term}). 
Calculations of higher order potential contain UV divergent terms. These divergences are canceled by introducing higher order terms
of $f$, and then we interpret the remaining finite 1-loop potential of
BL theory as the Newton potential.

Now we calculate $O(V^4)$ 1-loop potential with this $f$.  $O(V)$ term of the $f$ is sufficient to our purpose.  
For simplicity we consider Euclidean theory. Then the gauge fixed Lagrangian for massive fields with $f$ can be written as 
\begin{eqnarray}
\tilde{{\cal L}}=&&\frac{1}{2}\tilde{X}_{A}^{I}(\Box-g^2m_0^2)\tilde{X}^I_{A} \qquad (I=1, ..., 6, A=1,2) \nonumber\\
&+&\frac{1}{2}\bar{\Psi}^{1,2}(\partial \!\!\!/-\gamma^{78}gm_0)\Psi^{1,2}
+
\frac{1}{2}X^{\alpha}(\Box-g^2m_0^2+\delta m_X)X^{\alpha} 
\nonumber \\
&+&\frac{1}{2}A^{\alpha}(\Box-g^2m_0^2+\delta m_A)A^{\alpha}+HA^{\alpha}_{0}A^{\alpha}_{0}
+\bar{c}^{\alpha}(\Box-g^2m_0^2+\delta m_g+K\partial)c^{\alpha}
\nonumber \\
&+&J\epsilon^{ij}A^1_{i}A^2_{j}+F\epsilon^{ij}\partial_{i}A^{1,2}_{j}X^{1,2}\pm GA_0^{1,2}X^{2,1}.
\label{lagform}
\end{eqnarray}
Explicit forms of $m_0, \delta m, F,G,H,J$ are
\begin{eqnarray}
m_0^2&=&(L-\frac{V^2t^2}{L})^2, 
\hspace{3ex}
\delta m_{X}
=\frac{4V^2}{L^2}-\frac{\ddot\xi}{\xi}, 
\hspace{3ex}
\delta m_{A}=\frac{\ddot\xi}{\xi},
\nonumber \\
\delta m_g&=&\frac{\ddot f}{f}-\frac{2\dot{f}\dot{\xi}}{f\xi} ,
\hspace{3ex}
2H= 2\frac{\dot{\xi}^2}{\xi^2}-2\frac{\ddot\xi}{\xi},
\nonumber \\
K\partial&=&2\left(\frac{\dot{f}}{f}-\frac{\dot{\xi}}{\xi}\right)\partial_0, 
\hspace{3ex}
F=\frac{2V}{L\xi},
\hspace{3ex}
G=-\frac{4gV^2t}{L},
\hspace{3ex}
J=-\frac{4gV}{\xi^2}.
\end{eqnarray}
where $f=1\pm \frac{Vt}{L}$ in Euclidean theory. We define new fields from $A_{i}^{\alpha}$ as
\begin{eqnarray}
A_{1}^{1}=\frac{1}{\sqrt{2}}(\alpha_{2}+\beta_{1}), \hspace{1.7ex}
A_{2}^{1}=\frac{1}{\sqrt{2}}(-\alpha_{1}+\beta_{2}),  
\hspace{1.7ex}
A_{1}^{2}=\frac{1}{\sqrt{2}}(\alpha_{2}-\beta_{1}), \hspace{1.7ex}
A_{2}^{2}=\frac{1}{\sqrt{2}}(\alpha_{1}+\beta_{2}),
\end{eqnarray}
and   carry out Gaussian integration of $\tilde{X}^I_A$, $\Psi^{1,2}$, $X^{\alpha}$, $\bar{c}^{\alpha}, c^{\alpha}$ and $A^{\alpha}_0$. Then we obtain the following 1-loop effective Lagrangian 
:
\begin{eqnarray}
\tilde{{\cal L}}^{\rm 1 loop}=&-&6Tr\log(-\Box+m_0^2)
+4Tr\log(-\Box+m_0^2-\delta m_f)+4Tr\log(-\Box+m_0^2+\delta m_f) \nonumber \\
&-&Tr\log(-\Box_X)+2Tr\log(-\Box_g)-Tr\log(-\Box_{A_0})   
\nonumber \\
&+& \frac{1}{2}\alpha^{i}(\Box-m_0^2+\delta m_A+J)\alpha^i +\frac{1}{2}\beta^{i}(\Box-m_0^2+\delta m_A-J)\beta^i \nonumber \\
&-& \frac{1}{4} (\nabla\cdot \alpha -\nabla\times \beta) F\frac{1}{\Box_X}F (\nabla\cdot\alpha- \nabla\times \beta)
- \frac{1}{4} (\nabla\cdot \beta -\nabla\times \alpha) F\frac{1}{\Box_X}F (\nabla\cdot \beta- \nabla\times \alpha) \nonumber \\ 
&+&\frac{1}{16} (\nabla\cdot\alpha- \nabla\times \beta) F\frac{1}{\Box_X} 
G \frac{1}{\Box_{A_0}}G\frac{1}{\Box_X}F(\nabla\cdot\alpha -\nabla\times \beta)
\nonumber \\
&+&\frac{1}{16} (\nabla\cdot\beta- \nabla\times \alpha) F\frac{1}{\Box_X} 
G \frac{1}{\Box_{A_0}}G\frac{1}{\Box_X}F(\nabla\cdot\beta- \nabla\times\alpha ),
\label{efflag}
\end{eqnarray}
where the symbol $\Box$ is Laplacian and
\begin{eqnarray}
\delta m_f&=& \partial m_0=-\frac{2V^2t}{L},  \nonumber \\
\Box_X&=&\Box-m_0^2+\delta m_X,  \nonumber \\
\Box_g&=&\Box-m_0^2+\delta m_g+K\partial_t  \nonumber \\
\Box_{A_0}&=&\Box-m_0^2+\delta m_A+2H+G\frac{1}{\Delta_X}G.
\end{eqnarray}
We also introduce the notations $\nabla\cdot
\alpha=\partial_1\alpha_1+\partial_2\alpha_2$, $\nabla\times
\alpha=\partial_1\alpha_2-\partial_2\alpha_1$, and so on. In this
expression, we did not include contributions which  comes from massless
fields and tree level term $\frac{1}{2}(u^2+v^2)$.
Perturvative integration of $\alpha_i$, $\beta_i$ and expanding $\log$
determinants give the $O(V^4)$ 1-loop effective potential.
After a straightforward calculation we obtained
\begin{eqnarray}
-V^{\rm 1\;loop}(L,V)&=&\frac{1}{g\pi}\int d^3x\;\left[-\frac{V^4}{4L^5}+\frac{g^2V^4}{2L^3}t^2\right].
\end{eqnarray}
To have this results, we evaluated momentum integrals as follows:
\begin{align}
 &
  \int \! d^3x_1d^3x_2 ....d^3x_n\;\Delta(x_1,x_2)\Delta(x_2,x_3) .... 
  \Delta(x_{n-1},x_n)\Delta(x_n,x_1) =\int \! d^3x\;I(n), 
  \nonumber \\
 &
  \hspace{3ex}
  I(n)=\frac{\Gamma(n-\frac{3}{2})}{\Gamma(n)8\pi^{\frac{3}{2}}(m^2)^{n-\frac{3}{2}}}, 
  \quad 
  I(2)=\frac{1}{8\pi m}, 
  \quad 
  I(3)=\frac{1}{32\pi m^3}, 
  \quad 
  I(4)=\frac{1}{64\pi m^5}.
\\ 
 &\int \! d^3x_1 d^3x_2 \;t_1^2\Delta(x_1,x_2)\Delta(x_2,x_1)=\int d^3x\; t^2\;I(2).
\\ 
&\int \! d^3x_1 d^3x_2 \;t_1\Delta(x_1,x_2)t_2\Delta(x_2,x_1)=\int d^3x\;t^2\;J(2), \qquad 
J(2)=\frac{1}{128\pi^{\frac{1}{2}}\Gamma(\frac{3}{2})m} .
\end{align}
\begin{align}
&\int \! d^3x_1d^3x_2d^3x_3\;\left(\sum_{i=1,2}
 \frac{\partial}{\partial x^i_1}
 \frac{\partial}{\partial x_2^i}
 \Delta(x_1,x_2)\right)\Delta(x_2,x_3)\Delta(x_3,x_1)
=\int \! d^3x\;\frac{2}{3}[I(2)-m^3I(3)].
\\ 
&\int \! d^3x_1d^3x_2d^3x_3d^3x_4\;
\left(\sum_{i=1,2}\frac{\partial}{\partial x^i_1}\frac{\partial}{\partial x_2^i}\Delta(x_1,x_2)\right)\Delta(x_2,x_3)
\left(\sum_{j=1,2}\frac{\partial}{\partial x^j_3}\frac{\partial}{\partial x_4^j}\Delta(x_3,x_4)\right)\Delta(x_4,x_1) \nonumber 
\\
&=\int \! d^3x\;K(4), \qquad K(4)=\frac{1}{12\pi m},
\end{align}
where 
\begin{eqnarray*}
\Delta(x,y)=\frac{-1}{\Box-m^2}=\int\frac{dp^3}{(2\pi)^3}\frac{e^{ip(x-y)}}{p^2+m^2}
\end{eqnarray*} 
is the  free field  propagator with constant $m^2$.\\

Finally we discuss the higher order terms of $f$ beyond $O(V)$.
The 1-loop potential calculated by (\ref{efflag}) contains UV divergent contributions. The condition that these divergences are cancelled each other is 
\begin{eqnarray}
-\delta m_X+2\delta m_{g}-3\delta m_{A}-2H+\frac{2}{3}FF+\frac{1}{3}KK=0.
\label{nolindiv}
\end{eqnarray}               
It gives an equation to determine $f$:
\begin{eqnarray}
-\frac{4V}{L^2}+2\frac{\dot{\xi}^2}{\xi^2}+2\frac{\ddot{f}}{f}
-4\frac{\dot{f}\dot{\xi}}{f\xi}
+\frac{2}{3}\frac{4V^2}{L^2\xi^2}
+\frac{4}{3}\left(\frac{\dot{f}}{f}-\frac{\dot{\xi}}{\xi}\right)^2
=0.
\end{eqnarray} 
We introduce a new function $G=\frac{d}{dt}\log (f/\xi)$, then  obtain a differential equation 
\begin{eqnarray}
\dot{G}=-\frac{5}{3}G-\frac{d^2}{dt^2}\log \xi 
+\frac{2V^2}{L^2}\left(1-\frac{2}{3}\frac{1}{\xi^2}\right).
\label{difeq}
\end{eqnarray}
We can solve this equation order by order in $V$. Expand $G$ as $G=\sum^{\infty}_{n=1}G^{(n)}V^n$, then (\ref{difeq}) determines each $G^{(n)}$.
$G^{(1)}=\pm\frac{V}{L}$ to give no ( finite/infinite) correction to $O(V^2)$ terms of 1-loop potential.
In this way $f$ is determined as
 \begin{eqnarray}
f&=&\exp\left(\pm\frac{Vt}{L}-\frac{V^2t^2}{2L^2}\pm\frac{20V^3t^3}{9L^3}-\frac{115V^4t^4}{18L^4}+\cdots\right) \nonumber \\
&=&1\pm\frac{Vt}{L}+O(V^3).
\end{eqnarray}


\bigskip\bigskip


\end{document}